\documentstyle[aps,prl,epsf,preprint]{revtex}

\begin{document}  

\tightenlines
\newcommand{\ba}{\begin{eqnarray}}  
\newcommand{\ea}{\end{eqnarray}}  
\newcommand{\M}{{\cal M}}
\renewcommand{\S}{\Sigma}
\newcommand{\tc}{t_{+}}
\newcommand{\tO}{t_{-}}
\newcommand{\tf}{t_{0}}
\newcommand{\rc}{r_{+}}
\newcommand{\rO}{r_{-}}
\newcommand{\rf}{r_{0}}
\newcommand{\chic}{\chi_{+}}
\newcommand{\chio}{\chi_{-}}
\newcommand{\chif}{\chi_{0}}
\newcommand{\tk}{t_{(k)}}
\newcommand{\rk}{r_{(k)}}
\newcommand{\const}{{\it const.}}



\title{ 
\leftline{\baselineskip16pt\sl\vbox to0pt{
\hbox{\it Department of Physics}
\hbox{\it Osaka City  University}\vss}}
\rightline{\baselineskip16pt\rm\vbox to20pt{
\hbox{OCU-PHYS-181}
\hbox{~~~~~~~APGR-001}
\vss}}%
\vskip2cm
Big bang of the brane universe \\
}
\author{
   Hideki Ishihara\footnote{E-mail:~ishihara@sci.osaka-cu.ac.jp}
\vskip 10mm
}
\address{
        Department of Physics, Graduate School of Science, 
	Osaka City University, \\
	Sumiyoshi Osaka 558-8585, Japan  
}

\maketitle  

\begin{abstract}
Big bang of the Friedmann-Robertson-Walker (FRW)-brane universe is studied. 
In contrast to the spacelike initial singularity of the usual FRW universe, 
the initial singularity of the FRW-brane universe is point-like 
from the viewpoint of causality 
including gravitational waves propagating in the bulk. 
Existence of null singularities (\lq seam singuralities\rq) 
is also shown in the flat and open FRW-brane universe models. 
\end{abstract}



\section{Introduction}

It is widely accepted that the Friedmann-Robertson-Walker 
(FRW) universe model, which is the most simplified cosmological model, 
provides a successful account of the nature of the universe.
It is assumed, in this model, that we do not occupy a privileged 
position and do not find preferred directions in space. 
Then, a spatial section is considered as an isotropic 
and homogeneous space, which is 
described by a space of constant curvature. There are three 
possibilities: the spatial curvature is positive (closed model), 
zero (flat model) and negative (open model).
In all three cases the spatial sections has no boundary but 
a spatial volume is finite in the closed model and infinite 
in flat and open models. 

Expansion of the universe is described by time evolution of 
a scale factor that follows the Friedmann equation derived 
from the Einstein equation.  
If matter contained in the universe is approximated by homogeneous 
fluid with positive energy density and non-negative pressure, 
the universe began as the {\it big bang}, which is a singular 
state where the distance between all points of space was zero,
and the density of fluid and the curvature of spacetime were infinite.
Since the universe expands sufficiently rapidly from the big bang 
singularity, there exist particle horizons in the FRW models, 
in other words, the big bang singularity is spacelike 
from the viewpoint of causality.  

Recently, inspired by Horava-Witten theory\cite{Horava-Witten}, 
new attempts have been made 
on schemes of the $Z_2$-symmetric brane universe \cite{Lukas-etc} 
where matter fields are confined to a hypersurface embedded 
in a higher-dimensional spacetime, while only gravitational fields propagate 
through all of spacetime (see also\cite{old}). 
Randall and Sundrum\cite{Randall-Sundrum} 
have proposed a static vacuum brane solution, which is embedded 
in the five-dimensional anti-de Sitter bulk spacetime.
It is remarkable that the four-dimensional standard gravity 
on the brane is recovered with a 
small corrections in the low energy physics
\cite{Randall-Sundrum,Garriga-Tanaka,Shiromizu-Maeda-Sasaki}. 
The self-gravity of the brane plays an important role to single 
out the zero-mode, which corresponds to 4-dimensional gravity, 
as the ground state of the gravitational fields. 
Subsequently, many authors investigate on cosmological brane models 
which contains matter on the brane 
and derive the modified Friedmann equation\cite{cosmology,ida}. 

If the universe is a sub-spacetime embedded in a higher dimensional 
spacetime, there may be great changes of basic properties 
of the cosmological model. 
For example, it is pointed out that the causal structure of 
the closed FRW-brane universe 
is altered owing to the existence of gravitational waves 
propagating in the bulk\cite{Ishihara}. 
In this article, we study the geometrical structure of 
the FRW-brane universe 
at the initial stage in a framework of classical gravity,
and point out new properties of the singularities of the FRW-brane 
universe models. 


\section{Expansion of universe}
We consider a brane universe embedded in 
the five-dimensional anti-de Sitter spacetime. 
The five-dimensional anti-de Sitter spacetime can be decomposed 
locally as ${\cal N}_{(k)}^2\times{\cal M}_{(k)}^3$ where 
${\cal N}_{(k)}^2$ is a two-dimensional static spacetime and 
${\cal M}_{(k)}^3$ is a maximally symmetric 3-dimensional space 
characterized by a constant curvatures. 
The index $k$ takes one of three values, $+1,0,-1$, 
corresponding to the positive, zero and negative constant curvature 
of ${\cal M}_{(k)}^3$, respectively. 
The metric has a form 
\ba
  ds_5^2 = -f_{(k)}d\tk^2 + f_{(k)}^{-1} d\rk^2
			+ \rk^2 d\Omega_{(k)}^2,
\label{metric}
\ea
where
\ba
  f_{(k)}
	= \left\{
  \begin{array}{cl}
		1+\rc^2/l^2, 	\cr
		\rf^2/l^2, 			\cr
		\rO^2/l^2-1, 	\cr
  \end{array}
\right. \quad
	d\Omega_{(k)}^2 
	= \left\{
  \begin{array}{ll}\displaystyle
	 d\chic^2 + \sin^2{\chic}\{d\theta^2 + \sin^2{\theta}~d\phi^2\}
& \hspace{1cm}\mbox{for~~} k=+1,  \cr
	 d\chif^2 + \chif^2\{d\theta^2 + \sin^2{\theta}~d\phi^2\}
& \hspace{1cm}\mbox{for~~} k=0, \cr
	 d\chio^2 + \sinh^2{\chio}\{d\theta^2 + \sin^2{\theta}~d\phi^2\}
& \hspace{1cm}\mbox{for~~} k=-1 ,
  \end{array}
\right.
\label{omega}
\ea
respectively.
It might be noteworthy that $\rO$ is a time coordinate while $\tO$ is a space 
coordinate when $\rO<l$ in the case $k=-1$. 

Let us consider a embedding of a 4-dimensional hypersurface (three-brane), 
$\S$, with the structure of ${\cal N}^1\times{\cal M}_{(k)}^3$, 
where ${\cal N}^1$ is a trajectory of $\S$ in ${\cal N}_{(k)}^2$. 
We assume the position of ${\cal N}^1$ in ${\cal N}_{(k)}^2$ is given by
\ba
	r_{(k)}= a(\tau), \quad
	\tk= T(\tau),
\label{aT}
\ea
where $a$ and $T$ are functions of a parameter $\tau$. 
When $\tau$ is normalized by
\ba
  	-d\tau^2 =-f_{(k)}dT^2 + f_{(k)}^{-1} da^2, 
\label{proper-time}
\ea
the induced metric on $\S$ has a form of the Robertson-Walker metric
\ba
	ds^2 = -d\tau^2 + a^2(\tau)d\Omega_{(k)}^2,
\ea
where $a(\tau)$ plays a role of the scale factor of universe, 
and $k=+1,0,-1$ corresponds to the model with closed, flat and open 
spatial section, respectively. 

If matter is confined on $\S$ in the theory of brane, 
the surface energy-momentum tensor of the brane, $S_{ab}$, 
has a form of
\ba
	S_{ab} = -\sigma \gamma_{ab} + T_{ab}, \quad (a,b= 0,1,2,3), 
\label{stress}
\ea
where $\gamma_{ab}$ is the induced metric on $\S$, 
$\sigma$ is a constant which characterizes the brane tension,  
and $T_{ab}$ is the energy-momentum tensor of matter on $\S$. 
Since self-gravitating brane in the five-dimensional bulk should 
satisfy the metric junction condition\cite{Israel} with $Z_2$ symmetry, 
the extrinsic curvature of $\S$, $K_{ab}$, satisfies
\ba
	K_{ab} = -\kappa^2\frac12 ( S_{ab} - \frac13 S \gamma_{ab}), 
\label{junction}
\ea
where $\kappa^2$ is the five-dimensional gravitational constant. 

We assume, for simplicity, the energy-momentum tensor of matter 
takes the form of perfect fluid
\ba
	T_{ab} = (\rho+p) u_a u_b + p \gamma_{ab}, 
\label{perfect-fluid}
\ea
where $u^a$ is a 4-velocity field of the comoving observers on $\S$.

From (\ref{junction}) with (\ref{stress}) and (\ref{perfect-fluid}) 
we obtain the modified Friedmann equation\cite{cosmology,ida}
\ba
	\left( \frac{1}{a}\frac{d a}{d\tau}\right)^2
		&=& - \frac{k}{a^2} + \frac{8\pi}{3} \frac{\kappa^4\sigma}{6} \rho
			+ \frac{\kappa^4}{36}\rho^2,
\label{Hubble-eq}
\ea
and the conservation equation
\ba
	\frac{d}{d\tau}(\rho a^3) + p \frac{d}{d\tau}(a^3) =0, 
\label{cons-law}
\ea
where fine tuning $ 1/l^2 = \kappa^4 \sigma^2/36$ has been done 
so as to vanish the 4-dimensional effective cosmological constant 
on $\S$. 

If the equation of state is assumed as
\ba
	p= w\rho, \quad w=const.,
\ea
the conservation equation (\ref{cons-law}) implies that  
$
	\rho \propto a^{-q}$ with 
$q := 3(w+1)$,  
and then the Friedmann equation (\ref{Hubble-eq}) reduces to
\ba
	\left( \frac{1}{a}\frac{d a}{d\tau}\right)^2
		= - \frac{k}{a^2}
          + \frac{\alpha^2}{a^q}
		  + \frac{\beta^2}{a^{2q}}, 
\label{red-Hubble}
\ea
where $\alpha$ and $\beta$ are constants. 
If $q>1$ (equivalently $w>-2/3$), the last term in the right hand side 
of (\ref{red-Hubble}) dominates over other terms in the early stage 
where $a$ is small,  
then the asymptotic behavior of the scale factor as $\tau\rightarrow 0$ is 
\ba
	a(\tau) =  (4\beta\tau)^{1/q}.
\label{a->0}
\ea
The FRW-brane universe, in each case of $k=+1,0,-1$, 
begins from the singularity at $\tau = 0$ 
where the matter density diverges. 

For the cases $k=\pm1$, if $q>2$, 
the spatial curvature term, involving $k$ in the right hand side of
(\ref{red-Hubble}), becomes important in the late stage 
where $a$ is large. 
As the usual FRW models, the universe becomes a 
maximal size and then recollapses to the final singularity 
at a finite time after the big bang in $k=+1$ case, 
while the universe expands forever in $k=0$ and $k=-1$ cases. 
The asymptotic expansions of universe in large $\tau$ are 
same as the usual FRW models, 
{\it i.e.,} 
$\displaystyle \frac{da}{d\tau}\rightarrow 1$ as 
$\tau\rightarrow \infty$ in the case $k=-1$, 
while $\displaystyle \frac{da}{d\tau}\rightarrow 0$ 
as $\tau\rightarrow \infty$ in the case $k=0$.

On the other hand, from (\ref{proper-time}) we see the asymptotic 
behavior of $T(\tau)$ as $\tau \rightarrow 0$ in the form
\ba
	T(\tau) 
	= \left\{
  \begin{array}{cl}\displaystyle
	 (4\beta\tau)^{1/q}
& \hspace{1cm}\mbox{for~~} k=+1,  \cr
	 -(4\beta\tau)^{-1/q}
& \hspace{1cm}\mbox{for~~} k=0, \cr
	 (4\beta\tau)^{1/q}
& \hspace{1cm}\mbox{for~~} k=-1 .
  \end{array}
\right.
\label{T->0}
\ea
Then,  
$T \sim a \rightarrow 0$ as $\tau\rightarrow 0$ for $k=\pm 1$, while 
$T \sim -1/a \rightarrow -\infty$ as $\tau\rightarrow 0$ for $k=0$.
The difference of the behavior of $T(\tau)$ 
appears due to the fact that the metric component $f_\pm$ are regular at 
$\rk = 0$ but $f_0$ is singular.


\section{Charts}

Let us consider how the FRW-brane universe is embedded in 
the five-dimensional anti-de Sitter bulk spacetime. 
(See old suggestive works on embedding of 
the universe\cite{embedding}.) 
The five-dimensional anti-de Sitter spacetime is 
the universal covering space of the hyperboloid given by 
\ba
	-Y_0^2-Y_1^2+Y_2^2+Y_3^2+Y_4^2+Y_5^2= -l^2
\label{hyperboloid}
\ea
in the six-dimensional flat spacetime with the metric
\ba
	ds^2= -dY_0^2-dY_1^2+dY_2^2+dY_3^2+dY_4^2+dY_5^2 . 
\ea
The coordinates in (\ref{metric}) with (\ref{omega}) 
parameterize the hyperboloid in the following way;
\ba
\mbox{for}~ k=+1&&\cr
	&&Y_0= \sqrt{l^2+\rc^2}~ \sin{(\tc/l)}, \quad
	Y_1= \sqrt{l^2+\rc^2}~ \cos{(\tc/l)}, \cr
	&&Y_2= \rc \cos{\chic}, \quad
	Y_3= \rc \sin{\chic}~\cos{\theta}, 
\label{closed-chart}\\
	&&Y_4= \rc \sin{\chic}\sin{\theta}\cos{\phi}, \quad
	Y_5= \rc \sin{\chic}\sin{\theta}\sin{\phi}, \nonumber\\
\mbox{for}~ k=0~~&&\cr
	&&Y_2+Y_0= \rf, \quad
	Y_2-Y_0= \left(\frac{\tf^2}{l^2}-\chif^2\right)\rf-\frac{l^2}{\rf}, \cr
	&&Y_1= -\frac{\tf}{l}\rf, \quad
	Y_3= \rf\chif \cos{\theta}, 
\label{flat-chart}\\
	&&Y_4= \rf \chif \sin{\theta}\cos{\phi}, \quad
	Y_5= \rf\chif \sin{\theta}\sin{\phi}, \nonumber\\
\mbox{for}~ k=-1&&\cr
	&&Y_0= \rO \cosh{\chio}, \cr
	&&Y_1= \cases{
		\sqrt{l^2-\rO^2}~ \cosh{(\tO/l)} \cr
		\sqrt{\rO^2-l^2}~ \sinh{(\tO/l)} }, \quad
	Y_2= \cases{
		\sqrt{l^2-\rO^2}~ \sinh{(\tO/l)} \quad \mbox{for}~\rO<l   \cr
		\sqrt{\rO^2-l^2}~ \cosh{(\tO/l)} \quad \mbox{for}~\rO>l}, \cr
	&&Y_3= \rO \sinh{\chio}\cos{\theta}, \quad
	Y_4= \rO \sinh{\chio}\sin{\theta}\cos{\phi}, 
\label{open-chart}\\
	&&Y_5= \rO \sinh{\chio}\sin{\theta}\sin{\phi}. \nonumber
\ea
Substituting (\ref{aT}) 
into (\ref{closed-chart}), (\ref{flat-chart}),
and (\ref{open-chart}) for each case $k=+1,0,-1$,  
we get the four-dimensional embedded hypersurfaces, $\S$,  
parameterized by the cosmic time $\tau$ and comoving coordinates 
$\chi_{(k)}, \theta$ and $\phi$.

To see how the singularity $a=0$ is embedded, substitute 
(\ref{a->0}) and (\ref{T->0}) into (\ref{closed-chart}), 
(\ref{flat-chart}) and (\ref{open-chart}), respectively. 
Then, we get
\ba
	&&Y_1= l, \quad Y_0=Y_2= Y_3= Y_4= Y_5= 0 \qquad\qquad~ 
	\mbox{for ~}k=+1,
\label{closed-sing}\\
	&&Y_1= l, \quad Y_0+Y_2= 0, \quad
	Y_3= Y_4= Y_5= 0  \quad~~\quad \mbox{for ~}k=0,
\label{flat-sing}
\\
	&&Y_1= l, \quad Y_2= 0, \quad 
	-Y_0^2 + Y_3^2+Y_4^2+Y_5^2 = 0 \quad~ 
	\mbox{for ~}k=-1.
\label{open-sing}
\ea
It means that the singularity maps to a point for $k=+1$, 
a null line for $k=0$, and a three-dimensional null cone for $k=-1$.


\section{embedding}

We can better understand the embedding using 
the equation for $\S$ in the form of 
$F(Y_0,Y_1,Y_2,Y_3,Y_4,Y_5)=0$ by eliminating the 
parameter on the hypersurface. 
For elimination of $\tau$, 
it is rather easy to integrate a differential equation of $a$ with 
respect to $T$ directly. 
Using  (\ref{proper-time}) and (\ref{Hubble-eq}), 
we obtain
\ba
	\left(\frac{da}{dT}\right)^2 
	&=& f_{(k)}^2 \left[1 - f_{(k)}
  \left\{ l^{-2} a^2 + \alpha^2 a^{2-q}
		  + \beta^2 a^{2-2q}
		\right\}^{-1}\right]. 
\label{dadT}
\ea
Here, we concentrate on the initial behaviors of the 
FRW-brane universe. 
Since the last term in the curly bracket in (\ref{dadT}) dominates other 
two terms in the early stage, 
we get the initial behavior of $a$ up to next leading order:
\ba
 a \sim  \left\{
  \begin{array}{cl}\displaystyle
	l \tan\left(\frac{T}{l}
	  -\frac{l^6}{2(2q-1)\beta^2}\left(\frac{T}{l}\right)^{2q-1}\right)
&\quad \mbox{for}~ k=+1, \cr
\displaystyle
	 -l \left(\frac{l}{T} 
	-\frac{l^6}{2(2q-1)\beta^2} \left(\frac{T}{l}\right)^{-(2q-1)}\right)
&\quad \mbox{for} ~k=0, \cr
\displaystyle
	 l\tanh\left( \frac{T}{l} 
	  +\frac{l^6}{2(2q-1)\beta^2}\left(\frac{T}{l}\right)^{2q-1}\right)
&\quad \mbox{for} ~k=-1.
\end{array}
\right.
\label{a(T)}
\ea

Inserting $\rk = a(T)=a(\tk)$ into (\ref{closed-chart}), (\ref{flat-chart})
and (\ref{open-chart}), respectively, 
and eliminating the parameters $\chi_{(k)}, \theta$ and $\phi$, 
we obtain the equation of $\S$ in the form 
\ba
	&\displaystyle
	-l^2 \left(Y_0/Y_1\right)^2 +Y_2^2+Y_3^2 +Y_4^2+Y_5^2 
	=  -\frac{l^8}{(2q-1)\beta^2}\left(Y_0/Y_1\right)^{2q}
&\quad \mbox{for}~ k=+1, 
\label{closed-surface}\\ 
	&\displaystyle 
	-Y_0^2 +Y_2^2 +Y_3^2 +Y_4^2+Y_5^2 =
	 -\frac{l^8}{(2q-1)\beta^2} \left(Y_0/l+Y_2/l\right)^{2q}
&\quad \mbox{for}~ k=0, 
\label{flat-surface}
\\
	 &\displaystyle
	-Y_0^2 + Y_2^2 +Y_3^2 +Y_4^2+Y_5^2
		=  -\frac{l^8}{(2q-1)\beta^2}\left(Y_2/l\right)^{2q}
&\quad \mbox{for}~ k=-1.
\label{open-surface}
\ea

\begin{figure}[h]
  \begin{center}
    \leavevmode
    \epsfxsize=7cm
    \epsfbox{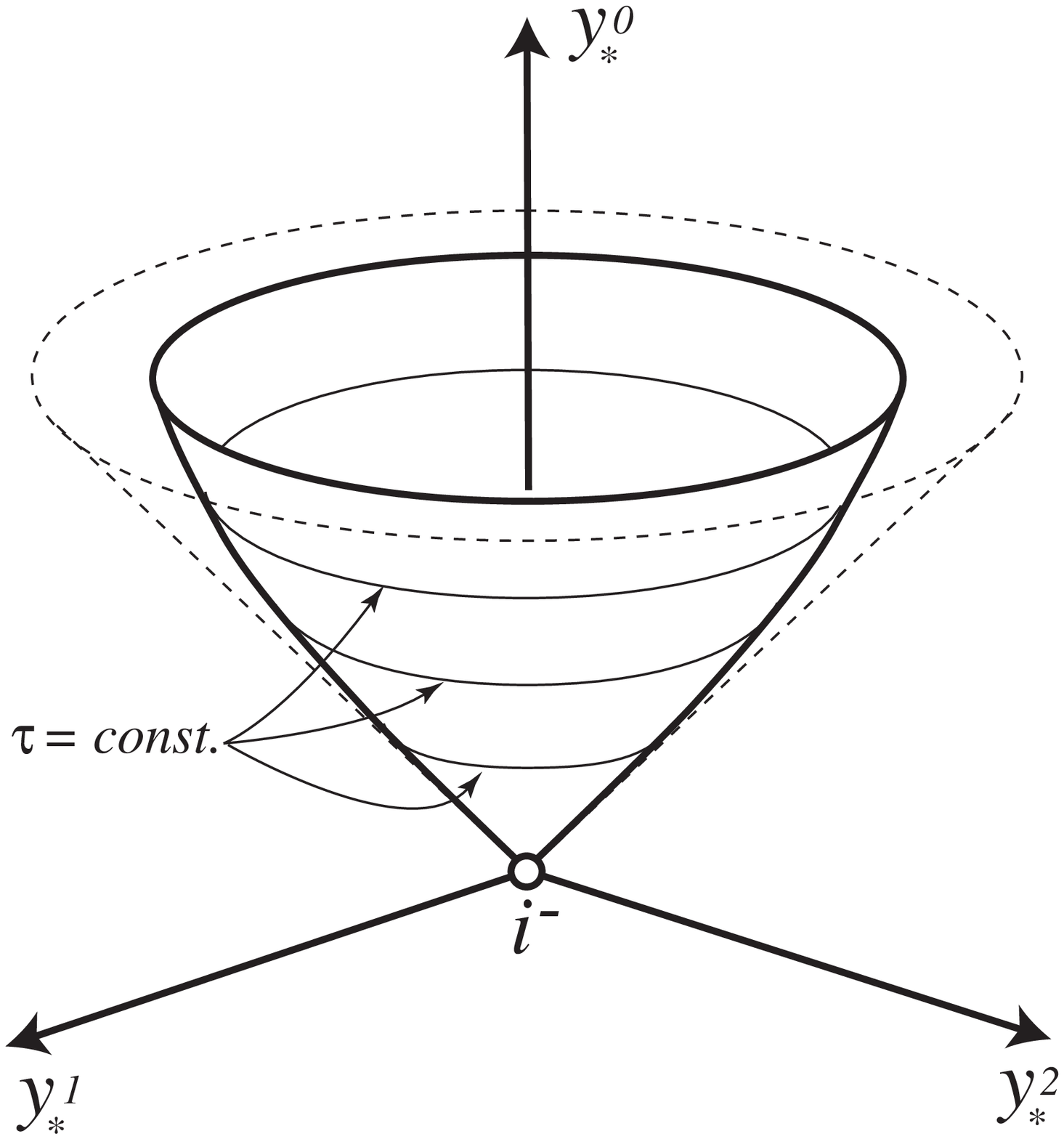}
    \caption{
Embedding of the closed FRW-brane universe at the initial stage 
with two coordinates, $y_*^3, y_*^4$, are suppressed. 
The vertex point $i^-$ denotes the initial big bang singularity. 
The null cone at $i^-$ generated by null geodesics in the bulk 
is drawn in broken line. 
}   
    \label{closed-brane}
  \end{center}
\end{figure}

\begin{figure}[h]
  \begin{center}
    \leavevmode
    \epsfxsize=7cm
    \epsfbox{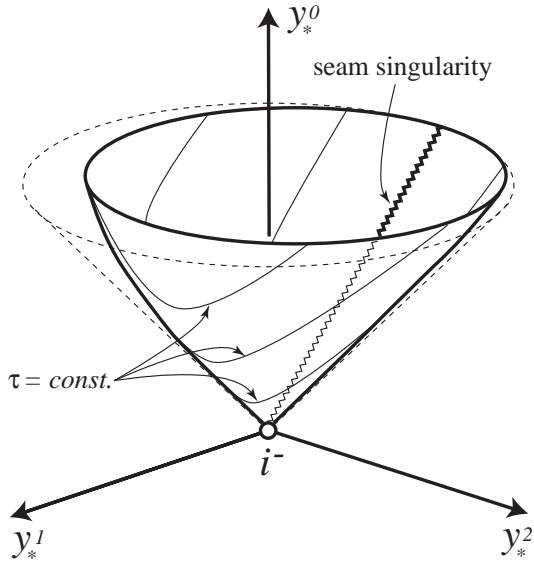}
    \caption{
Embedding of the flat FRW-brane universe at the initial stage. 
The seam singularity lies on the brane. 
}   
    \label{flat-brane}
  \end{center}
\end{figure}

\begin{figure}[h]
  \begin{center}
    \leavevmode
    \epsfxsize=7cm
    \epsfbox{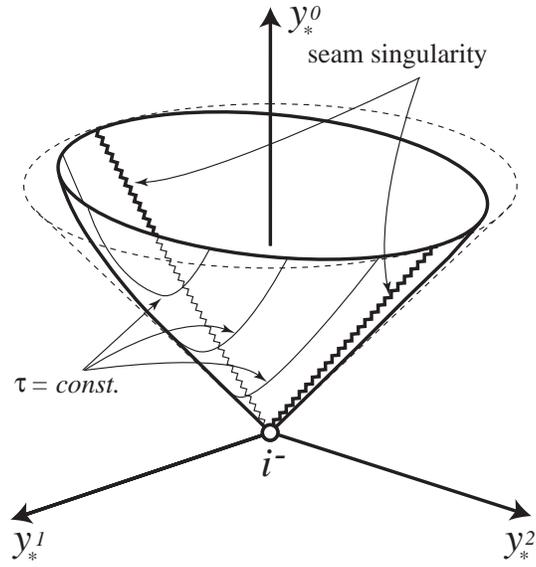}
    \caption{
Embedding of the open FRW-brane universe at the initial stage. 
}   
    \label{open-brane}
  \end{center}
\end{figure}

The coordinates $Y_A$ are not convenient to visualize how the brane is 
embedded because they have a redundant component which can be 
eliminated by the constraint (\ref{hyperboloid}).
We use, instead, the closed chart (\ref{closed-chart}) 
because it covers whole anti-de Sitter space time regularly. 
Furthermore, for convenience to discuss the causal structure 
we introduce a new radial coordinate that makes the metric (\ref{metric}) 
has a conformally flat form in the temporal-radial subspace in the form
\ba
  ds_5^2
	&=&f_+ \left( - d\tc^2 +  dr_*^2 \right)
		+ \rc^2 d\Omega_+^2.
\ea
By use of $r_*$ we define new coordinates as follows;
\ba
	 &&y_*^0 :=\tc, \quad y_*^1 := r_* \cos\chic, \quad 
	y_*^2 := r_* \sin\chic\cos\theta, \cr
	 &&y_*^3 := r_* \sin\chic\sin\theta\cos\phi, \quad 
	y_*^4 := r_* \sin\chic\sin\theta\sin\phi.
\label{newcoord}
\ea
In this coordinate system, radial null geodesics are represented by 
straight lines at $\pm 45^\circ$. 
Using (\ref{newcoord}) we rewrite 
(\ref{closed-surface})-(\ref{open-surface}) as
\ba
	&\displaystyle 
	-(y_*^0)^2 + (y_*^1)^2+(y_*^2)^2+(y_*^3)^2+(y_*^4)^2 
		= -\frac{l^8}{(2q-1)\beta^2}(y_*^0/l)^{2q} 
&\quad \mbox{for}~ k=+1, 
\label{closed-emb}\\
	&\displaystyle 
	-(y_*^0)^2 + (y_*^1)^2+(y_*^2)^2+(y_*^3)^2+(y_*^4)^2 
		= -\frac{l^8}{(2q-1)\beta^2}(y_*^0/l+y_*^1/l)^{2q} 
&\quad \mbox{for}~ k=0,  
\label{flat-emb}\\
	&\displaystyle 
	-(y_*^0)^2 + (y_*^1)^2+(y_*^2)^2+(y_*^3)^2+(y_*^4)^2 
		= -\frac{l^8}{(2q-1)\beta^2}(y_*^1/l)^{2q}
&\quad \mbox{for}~ k=-1. 
\label{open-emb}
\ea
Embedded hypersurfaces in $y_*^0, y_*^1, y_*^2$ spacetime 
are shown in Figs.\ref{closed-brane}, \ref{flat-brane} and \ref{open-brane}, 
where $y_*^3, y_*^4$ are suppressed by setting $\theta = 0, \pi$ 
since the directions $y_*^2, y_*^3, y_*^4$ are isotropic.

It is clear that the brane in each case is smooth everywhere 
except the vertex point, $i^-$, 
located on $y_*^a=0 (a=0,1,2,3,4)$. 
Future directed null geodesics emanating 
from $i^-$ form a null cone which is drawn by broken lines in the figures, 
and the brane is tangent to the null cone at $i^-$.
In the case $k=+1$, $\S$ has a shape of conoid 
which is isotropic in $y_*^1, y_*^2, y_*^3, y_*^4$ directions.  
The hypersurface is timelike everywhere except $i^-$. 
On the other hand, in the case $k=0$, $\S$ is a conoid 
which is tilted towards $-y_*^1$ 
direction and deformed so that a null line 
\ba
	y_*^0 + y_*^1=0, \quad y_*^2=y_*^3=y_*^4=0
\label{seam-flat}
\ea
lies on $\S$. 
Since the induced metric on $\S$ degenerates there, 
then this line is a singular subspace on the brane universe. 
It seems that one sheet of brane is curled up 
and seamed two edges of the brane together 
along the null line (\ref{seam-flat}). 
We call the line, therefore, \lq seam singularity\rq. 
Similarly, in the case $k=-1$, $\S$ is a conoid which is squashed 
in $\pm y_*^1$ directions so that a null cone 
\ba
	-(y_*^0)^2 +(y_*^2)^2+(y_*^3)^2+(y_*^4)^2=0, \quad y_*^1=0 
\label{seam-open}
\ea
lies on $\S$. In Fig.\ref{open-brane} two seams represent the 
cross section of the null cone (\ref{seam-open}) 
and the hyperplane $y_*^3=y_*^4=0$. 
In this case, the seam singularity is not a line but 
a three-dimensional surface which divides $\S$ into two parts.

We construct the total spacetime as follows; 
cut the five-dimensional anti-de Sitter spacetime along $\S$, 
keep the side surrounded by $\S$, make a copy of the part 
of five-dimensional anti-de Sitter spacetime with the boundary $\S$, 
join these two with identifying the both boundaries $\S$.
It should be stressed that the past region in total spacetime 
about $i^-$ does not exist, {\it i.e.}, 
the brane and bulk begin simultaneously at the 
initial big bang singularity, $i^-$. 

Cosmic time slices $\tau=\const$ of the FRW-brane universe coincide with 
the restriction of $y_*^0=\const$ on $\S$ for $k=+1$, 
with  $y_*^0+y_*^1 =\const$ for $k=0$, and 
with $y_*^1 =\const$ for $k=-1$, respectively. 
For $k=0$ and $k=-1$ cases, the slices can be extended to 
infinity without intersecting the seam singularity. 
Then the spatial volume of the universe along the slice is infinite. 
The constant $\tau$ surfaces, which respect the Killing vector fields 
on the brane, have a nice feature, {\it i.e.}, 
they are constant density surfaces. 
However, it is also possible to take another time slice which 
intersects the seam singularity, for example, 
the restriction of $y_*^0=\const$ on $\S$. 
If we take this time slices the spatial distance measured along a curve 
lies on the slice from a point on $\S$ to the seam singularity is finite, 
{\it i.e.,} 
the spatial volume is finite and the edge of the spatial 
section of universe is the seam singularity. 

All of comoving coordinates constant lines 
$\chi_{(k)}=\const, \theta=\const,  \phi=\const$, 
which are perpendicular to the constant $\tau$ surfaces, 
converge backward in time to $i^-$. 
Thus, this point is the initial big bang singularity of the universe. 
The initial singularity is point like in all of closed, flat and 
open FRW-brane models 
in contrast to the fact that the initial singularity is space like 
in the usual FRW models. 
Therefore, there is no particle horizon in the FRW-brane models.

We can consider a tangent space about an arbitrary point on the 
seam singularity because $\S$ is smooth there. 
Since the seam singularity is null subspace, the tangent space of $\S$ 
at a point on the seam singularity is spanned by 
a null vector and spacelike vectors. 
Any causal curve confined on $\S$ 
which starts from the point on the seam singularity 
is a null geodesic, which is a generator of the seam singularity 
itself. 
Therefore, there is no causal curve on $\S$ which connects 
a regular point on $\S$ and the seam singularity.
It means that the seam singularity is invisible by use of physical 
phenomena confined on the brane.
In other words, the seam singularity is spatial infinity 
with respect to the causality restricted on the brane. 
However, there exists a causal curve penetrating the bulk 
which connects a point on $\S$ and the seam singularity.
It means that the seam singularity is visible 
by gravitational waves propagating in the bulk.

In the case of flat and open FRW-brane models, 
let us take two points on a $\tau = \const$ surface and 
consider affine distances from these point to the seam singularity 
along null geodesics of gravitons with same frequencies in the bulk. 
It is obvious that these distances take different values, generally.
Then, these tow points, which lie on an isotropic and homogeneous 
surface, are equivalent in the intrinsic geometry, 
but these are not equivalent in the position relative to the seam singularity. 
In contrast, any two points on a $\tau = \const$ surface in the closed 
model are equivalent both in the brane and in the bulk. 

The consideration in the paper is based on the classical 
description of the brane which might be modified in the 
early stage of the universe. 
It is interesting problem to clarify the global structure 
of the brane universe including quantum effects.


\bigskip

\begin{acknowledgements}
The author would like to thank N.Deruelle for her encouragement. 
He gratefully acknowledges the hospitality of DARC, Paris Observatory, 
Meudon, where this work was initiated. 
\end{acknowledgements}




%


%

\end{document}